\documentclass[twoside,twocolumn,9pt]{article}
\usepackage{extsizes}
\usepackage[super,sort&compress,comma]{natbib} 
\usepackage[left=1.5cm, right=1.5cm, top=1.785cm, bottom=2.0cm]{geometry}
\usepackage{balance}
\usepackage{sectsty}
\usepackage{graphicx} 
\usepackage{lastpage}
\usepackage{float}
\usepackage{fancyhdr}
\usepackage{fnpos}
\usepackage[english]{babel}
\addto{\captionsenglish}{%
  
}
\usepackage{array}
\usepackage[symbol]{footmisc}

\usepackage{charter}
\usepackage[T1]{fontenc}
\usepackage[usenames,dvipsnames]{xcolor}
\usepackage{setspace}
\usepackage[compact]{titlesec}
\usepackage{hyperref}
\usepackage{bm}
\definecolor{cream}{RGB}{222,217,201}
\usepackage{changes}

\usepackage{amsmath}
\usepackage{amsfonts}
\usepackage{amssymb}
\usepackage{todonotes}
\usepackage{float}
\def\bea{\begin{eqnarray}}
\def\eea{\end{eqnarray}}
\def\la{\langle}
\def\ra{\rangle}

\def\bl{\begin{align}}
\def\el{\end{align}}

\newcommand{\eref}[1]{eqn~(\ref{#1})}%
\newcommand{\Fref}[1]{Figure~\ref{#1}}%

\begin{document}

\pagestyle{fancy}
\thispagestyle{plain}


\twocolumn[
  \begin{@twocolumnfalse}

{\huge\textbf{Direction reversing active Brownian particle in a harmonic potential}}

 \noindent{\Large Ion Santra,\textit{$^{a}$} Urna Basu,\textit{$^{a,b}$} and Sanjib Sabhapandit\textit{$^{a}$}} \\
 
 \noindent $^a$ {\it Raman Research Institute, Bengaluru 560080, India}\\
\noindent $^b$ {\it S. N. Bose National Centre for Basic Sciences, Kolkata 700106, India.}\\

{\bf Abstract}\\
\noindent We study the two-dimensional motion of an active Brownian particle of speed $v_0$, with intermittent directional reversals in the presence of a harmonic trap of strength $\mu$. The presence of the trap ensures that the position of the particle eventually reaches a steady state where it is bounded within a circular region of radius $v_0/\mu$, centered at the minimum of the trap. Due to the interplay between the rotational diffusion constant $D_R$, reversal rate $\gamma$, and the trap strength $\mu$, the steady state distribution shows four different types of shapes, which we refer to as active-I \& II, and passive-I \& II phases.
In the active-I phase, the weight of the distribution is concentrated along an annular region close to the circular boundary, whereas in active-II, an additional central diverging peak appears giving rise to a Mexican hat-like shape of the distribution. The passive-I is marked by a single Boltzmann-like centrally peaked distribution in the large $D_R$ limit. On the other hand, while the passive-II phase also shows a single central peak, it is distinguished from passive-I by a non-Boltzmann like divergence near the origin. We characterize these phases by calculating the exact analytical forms of the distributions in various limiting cases. In particular, we show that for $D_R\ll\gamma$, the shape transition of the two-dimensional position distribution from active-II to passive-II occurs at $\mu=\gamma$. We compliment these analytical results with numerical simulations beyond the limiting cases and obtain a qualitative phase diagram in the $(D_R,\gamma,\mu^{-1})$ space.

\end{@twocolumnfalse} \vspace{0.6cm}

  ]


\section{Introduction}

Brownian motion is perhaps the simplest stochastic process that has found diverse applications across a wide range of disciplines including natural sciences~\cite{brown1,brown2}, ecology~\cite{eco}, computer sciences~\cite{comp} and finance~\cite{finance}. The paradigmatic example is the jittery motion of a  micron-sized colloidal particle in a fluid at a temperature $T$. The dynamics of the position vector $\bm{r}$ of such a \emph{passive} Brownian particle in the presence of a confining potential $V(\bm{r})$ is described by the overdamped  Langevin equation $\dot{\bm {r}}=-\alpha \nabla V(\bm {r})+\sqrt{2\alpha k_BT}\,\bm {\eta}(t)$, where $\alpha$ is the mobility, $k_B$ is the Boltzmann constant, and $\bm{\eta}(t)$ is a delta-correlated Gaussian white noise. The position of the particle eventually equilibrates to the Boltzmann distribution $\propto \exp\left[- V(\bm {r})/(k_BT)\right]$. 

A bacterium, like {\it E. coli}, which has a size similar to a colloidal particle, performs, on the other hand, a very different kind of motion~\cite{berg1}. It self-propels with a constant speed $v_0$ along an internal orientation vector $\bm{\hat{n}}$, that itself evolves stochastically. Such directed/persistent motion $\dot{\bm {r}}(t)=v_0\bm{\hat{n}}(t)$, referred to as active motion, have gained a lot of interest in recent times~\cite{abp1,hydrodynamics1,roadmapactive,Sriram,Bechinger}.
Active Brownian particle (ABP)~\cite{abp0,cates_abprtp,separation_abp,abp_potoski,franosch1} and run-and-tumble particle (RTP)~\cite{rtp2008,abprtp_comp,pressure,rtpddim} are two widely used models to describe different kinds of active motion. The orientation vector $\bm{\hat{n}}$ undergoes a rotational diffusion for ABP~\cite{sevilla2014,abp2018}, while in the case of RTP, intermittent \emph{tumbling} events results in the reorientation of $\hat{n}$ along a randomly chosen direction~\cite{rtp2d_2012,ion1}.

In the presence of a confining potential, an active particle relaxes to a nonequilibrium stationary state, whose form depends on the potential and the specific dynamics of $\bm{\hat{n}}$~\cite{Bechinger} --- unlike the generic equilibrium Boltzmann distribution for the passive case. There have been a handful of theoretical studies that find exact results for the stationary state in such scenarios~\cite{abp_potoski,abp2019,abp2020,1drtptrap}. It turns out that the presence of activity leads to a wide range of non-trivial behaviors. Of particular interest is the shape-transition of the position distribution from an \emph{active phase}, characterized by an accumulation of probability density near the boundary of the confining region, to a Boltzmann-like \emph{passive phase}~\cite{abp_potoski,abpexp,1drtptrap,abp2018}. Naturally, exploring the stationary-state behavior of various active motions in confining potentials is of significant interest.

Certain bacteria like {\it Myxococcus xanthus}~\cite{xanthus1,xanthus2,xanthus3,xanthus4}, {\it Pseudomonas putida}~\cite{putida1,putida2}, {\it Pseudoalteromonas haloplanktis and Shewanella putrefaciens}~\cite{marine1,marine2}, and {\it Pseudomonas
citronellolis}~\cite{monoperitrichous} show a unique type of motion, not described by either ABP or RTP. They undergo ABP like motion accompanied by intermittent reversals of the orientation vector. Various theoretical models for such motion have been explored recently~\cite{drabp,grossman1,detcheverry}. In particular, when the reversal events follow a Poisson process with a constant rate, such a direction reversing active Brownian particle (DRABP) shows non-trivial position distribution and persistence properties in the absence of any external potential~\cite{drabp}. A natural direction is to investigate the steady-state behavior of a DRABP in confining potentials.

\begin{figure*}[h!]
\centering{
\includegraphics[width=0.9\hsize]{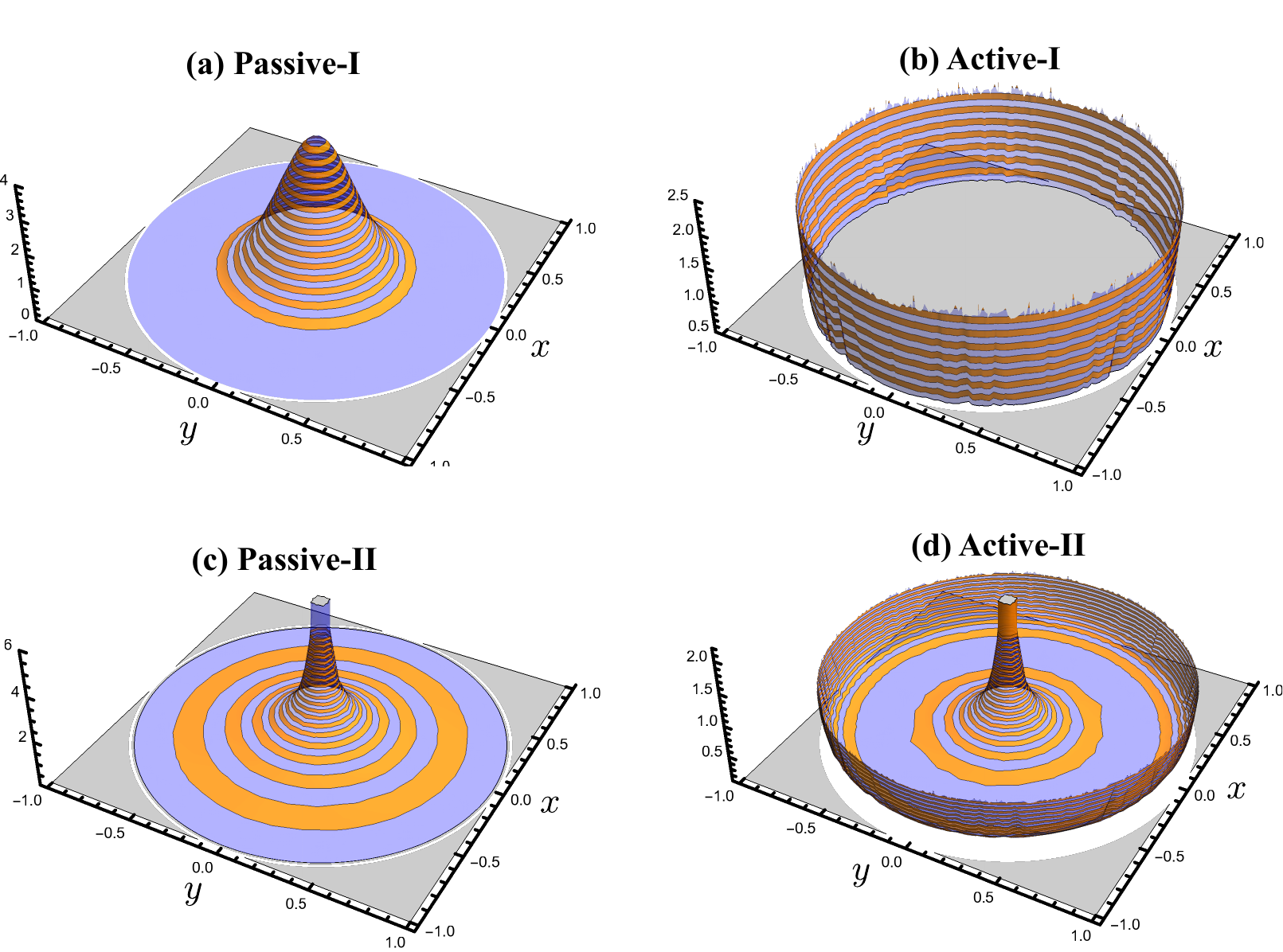}}
\caption{Typical shape of the stationary position distribution $P(x,y)$ for $r_0=1$ in the four phases---(a) Passive-I [\eref{gaus_drabpXY}], (b) Active-I [\eref{eq:gsmall1}], (c) Passive-II [\eref{eq:2dsc} for $\nu>1$], (d) Active-II  [\eref{eq:2dsc} for $\nu<1$].}
\label{2dplot}
\end{figure*}

In this paper, we study the stationary position distribution of a DRABP in two dimensions, in the presence of a harmonic potential. We show that the interplay between the rotational diffusion, direction reversal and the harmonic trap leads to four phases characterized by distinct shapes of the position distribution (see Fig.~\ref{phasediagram} and \ref{2dplot}). Apart from the typical active and passive phases---marked by an accumulation of probability at the boundaries and a Boltzmann-like centrally peaked distribution respectively---we find two novel phases where a diverging central  peak appears in both active and passive phases. We characterize the transition/crossover among these phases.

The paper is organized as follows. We define the model and announce our main results along with a qualitative phase diagram in Sec.~\ref{sec:model}. In Sec.~\ref{sec:trajectory} we provide a qualitative description of the different phases in terms of long-time trajectories. Detailed analytical derivations of the position distributions in the different phases are provided in Sec.~\ref{sec:ss_dist}. Finally, we conclude in Sec.~\ref{sec:concl} with some open questions. Exact computation of the variance and kurtosis is given in Appendix~\ref{sec:moment}. We also generalize part of our results to arbitrary dimensions in Appendix~\ref{sec:ddim}.

\section{The model and results}\label{sec:model}

 A direction reversing active Brownian particle moving in two dimensions is described by its position vector $\bm{r}=(x,y)$, orientation angle $\theta\in[0,2\pi]$ with respect to the $x$-axis, and the dichotomous noise $\sigma(t)=\pm1$. In the presence of a harmonic potential
\begin{equation}
  V(x,y)=\frac {\mu}{2} (x^2+y^2),
  \label{potential}
\end{equation}
   the position and orientation evolve according to the Langevin equations,
\begin{subequations}
\label{langevin}
\begin{align}
\label{eq:x}
\dot{x}&=-\mu x+v_0\,\sigma (t)\, \cos\theta (t),\\
\label{eq:y}
 \dot{y}&=-\mu y+v_0\,\sigma (t)\, \sin\theta (t),\\ 
 \label{eq:theta}
 \dot{\theta}&= \sqrt{2D_R}\,\eta(t).
\end{align}
\end{subequations}
Here $D_R$ is the rotational diffusion constant and $\eta(t)$ is a Gaussian white noise with $\la \eta(t)\ra=0$ and $\la\eta(t)\eta(t')\ra=\delta(t-t')$. The dichotomous noise $\sigma(t)$ flips between $\pm 1$ at a constant rate $\gamma,$ triggering orientation reversals. It has an exponentially decaying autocorrelation $\la \sigma(t)\sigma(t')\ra=e^{-2\gamma|t-t'|}$. 

For $\gamma=0$, this model reduces to the ABP in a harmonic potential, for which the stationary state has been studied in~\cite{abp2018,abp2020}. On the other hand, for $D_R=0$, since $\theta$ does not evolve, the model corresponds to a one-dimensional RTP along the initial orientation in a  harmonic potential~\cite{1drtptrap}. In the absence of any potential, for both $\gamma=0$ (ABP) and $D_R=0$ (RTP), the long-time dynamics becomes diffusive with effective diffusion coefficient $D_{\text{AB}}=v_0^2/(2D_R)$ and $D_{\text{RT}}=v_0^2/(2\gamma)$, respectively. For DRABP, i.e., both non-zero $\gamma$ and $D_R$, the corresponding effective diffusion coefficient is $D_{\text{DR}}=v_0^2/[2(D_R+2\gamma)]$. In this paper, we find that, in the presence of a harmonic potential, the interplay of $\gamma$, $D_R$ and $\mu$ leads to a host of interesting behaviors in the stationary state.

The Fokker-Planck equation for the  probability density function $P_{\sigma}(x,y,\theta,t)$ corresponding to the Langevin equations~\eqref{langevin} is given by,
\begin{align}
\frac{\partial P_{\sigma}}{\partial t}=&-\left[\frac{\partial}{\partial x}(-\mu x+v_0\sigma\cos\theta)+\frac{\partial}{\partial y}(-\mu y+v_0\sigma\sin\theta)\right]P_{\sigma}\nonumber\\ &-\gamma\, P_{\sigma}+\gamma\, P_{-\sigma}+D_R\frac{\partial^2P_{\sigma}}{\partial \theta^2}.
\label{eq:fp}
\end{align}
We are interested in the steady state position distribution
 \begin{align}
P(x,y)\equiv\int_{0}^{2\pi}d\theta \sum_{\sigma=\pm1}P_{\sigma}(x,y,\theta),
\end{align}
  where the stationary distribution $P_{\sigma}(x,y,\theta)\equiv P_{\sigma}(x,y,\theta,t\to\infty)$ is the  solution of \eref{eq:fp} with $\partial P_{\sigma}/\partial t=0$. \footnote[2]{Note that, for notational simplicity, we are using the same letter $P$ to denote all the probability distributions.} The exact solution of \eref{eq:fp} is hard to obtain in practice, for arbitrary values of $\mu,\,\gamma$ and $D_R$, even for the steady state. Hence we analyze the distribution $P(x,y)$ in the limiting cases where one of the parameters is much smaller than the others, giving rise to distinct phases characterized by the shape of the position distribution. It is evident from \eref{langevin} that the steady state is isotropic and has a finite support on a circular region  of radius $r_0=v_0/\mu$ centered at the origin. We show that, depending on the relative strength of the three parameters $D_R$, $\gamma,$ and $\mu$, the shape of the steady state position distribution can be very different (see Fig.~\ref{2dplot}), which we analytically characterize. 

Before going to the detailed analysis, we introduce the different phases and briefly summarize our main results here. We analytically find that for $\gamma\ll\mu$ or $D_R\ll\mu$, the system is in an \emph{active phase}, where the probability density accumulates near the circular boundary of radius $r_0$. On the other hand, for  $\gamma\gg\mu$ or $D_R\gg\mu$ the system is in a \emph{passive phase} where the distribution has a single central peak.
A unique feature of this DRABP in harmonic potential is that for $ D_R\ll\gamma$, the distribution at the center always diverges (algebraically for the two-dimensional position distribution and logarithmically for the marginal) irrespective of whether the system is in the active or the passive phase. To take this into account, we further subdivide the each of two phases into two sub-phases based on our analytical results in the limiting cases.

\begin{figure*}[htp]
\centering{
\includegraphics[scale=0.9]{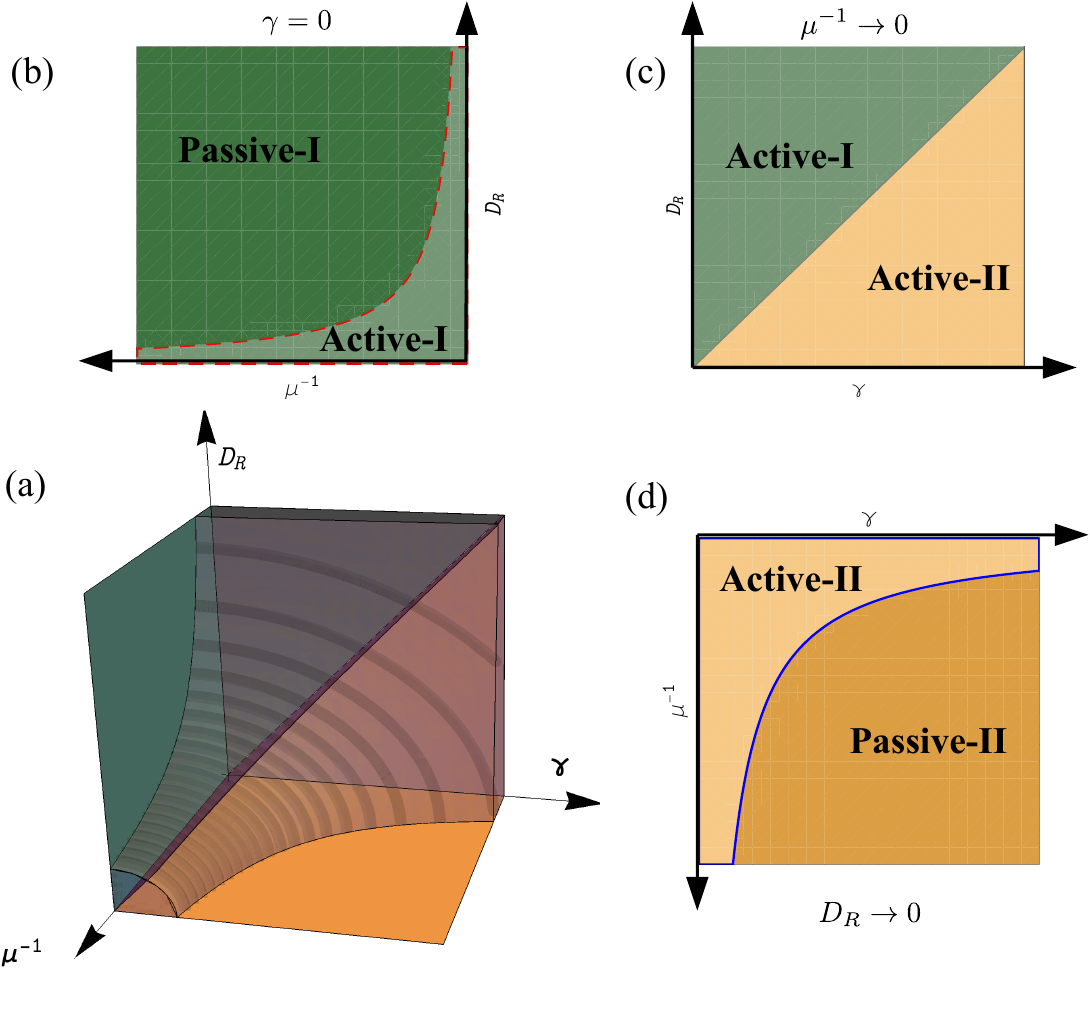}}
\caption{A simplistic schematic phase diagram for the steady state of DRABP in a harmonic trap. (a) The phase diagram in the $(\gamma,\,D_R,\,\mu^{-1})$ space. For $\gamma= 0$, the phase diagram of DRABP becomes that of an ABP in a harmonic trap [shown separately in (b)], where  we see the active-I phase for $D_R\ll\mu$ (light green shaded region), which crosses over to the passive-I phase for $D_R\gg\mu$ (dark green shaded region). Although we do not know the crossover/transition curve analytically, for simplicity, it is  shown by the schematic dotted red line. The $D_R\to 0$ plane [shown separately in (d)] shows a transition from the active-II to passive-II phase, where the transition line, marked by a solid blue line, is known exactly (see Sec.~\ref{sec:dr0}).  Numerical evidence suggests (see Sec.~\ref{numericalss}) that the behaviors shown in (b) and (c) extend for $\gamma>0$ and $D_R>0$ respectively, implying some kind of transition/crossover between active-I and active-II as well as passive-I and passive-II phases. The active (I \& II) region shrinks as $\mu^{-1}$ increases, indicating the funnel-like shape of the surface separating the active (I \& II) and the passive (I \& II) phases shown in (a). However, the actual shape of this surface may have more complex structure [see  Fig.~\ref{f:u}(b)] than the simple schematic surface shown here. For $\mu^{-1}\to 0$ [shown separately in (c)], the passive region disappears (i.e., pushed to infinity) leaving only the active phases.
}
\label{phasediagram}
\end{figure*}

\begin{itemize}
\item \textbf{Passive-I} ($D_R\gg\mu$, for arbitrary $\gamma$). In this case, the stationary distribution is Boltzmann-like which has a Gaussian form for the harmonic potential considered here [see Fig.~\ref{2dplot}(a)]. This is similar to the typical passive phase seen for ABP ($\gamma=0$) in an external potential.

\item \textbf{Active-I} ($\gamma\ll D_R\ll\mu$).  Here the stationary distribution is concentrated at the circular boundary $|\bm{r}|=r_0$ [see Fig.~\ref{2dplot}(b)]. This is also the active phase for ABP, where $\gamma=0$.

\item \textbf{Passive-II} ($\gamma>\mu\gg D_R$). In this passive phase also, the position distribution has a single central peak. However, the distribution diverges at the center which distinguishes it from the passive-I phase [see Fig.~\ref{2dplot}(c)].

\item \textbf{Active-II} ($\mu>\gamma\gg D_R$). This phase is characterized by a Mexican hat-like shape [see Fig.~\ref{2dplot}(d)] of the distribution that is concentrated both at the center and at the circular boundary $|\bm{r}|=r_0$.

\end{itemize}
While we have characterized the above phases analytically only in the limiting cases, the general qualitative features hold even beyond these limits, which we verify using numerical simulations for some other parameters (see Fig.~\ref{f:u}). The phases are best represented in the $\gamma,$ $D_R$ and $\mu^{-1}$ space and a qualitative phase diagram is provided in Fig.~\ref{phasediagram}. In order to develop a physical understanding of the emergence of the different shapes, we look at the typical trajectories in the different phases in the following section.

\section{Typical trajectories in the different phases}\label{sec:trajectory}

\begin{figure}
\centering{
\includegraphics[width=0.9\hsize]{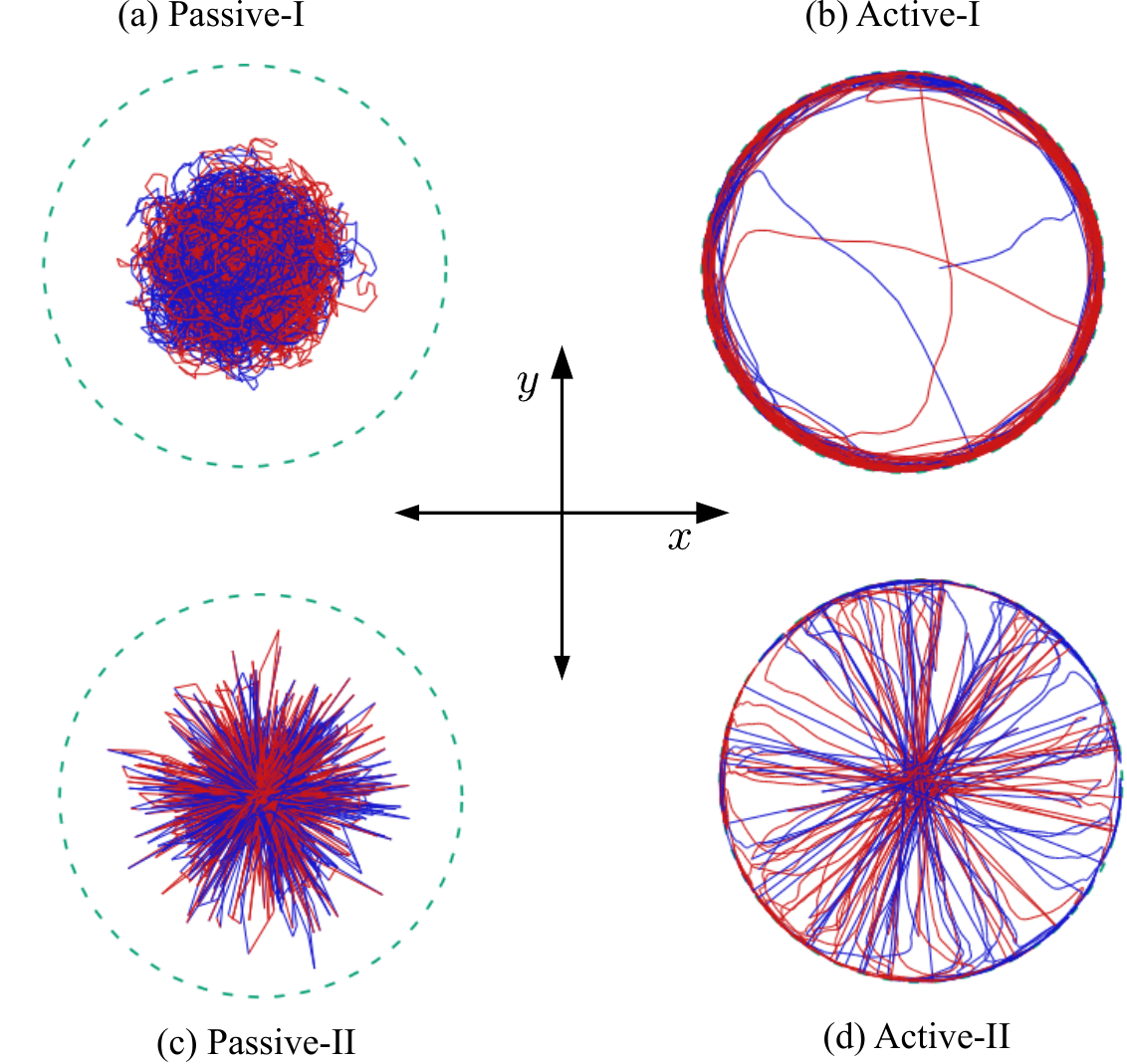}}
\caption{Long-time trajectories of a DRABP in a harmonic trap with $\mu=1$ in the different phases. The finite support of the position distribution, a circle of radius $r_0=1$, is marked by dashed green lines. (a) passive-I phase: $\gamma=0.1$ $D_R=10$; (b) active-I phase: $\gamma=0.001$ $D_R=0.05$ (c) passive-I phase: $\gamma=5$ $D_R=0.01$ (d) active-II phase: $\gamma=0.1$ $D_R=0.01$. The blue and red colors indicate the instantaneous state $\sigma=+1$ and $-1$ respectively.}
\label{f:trajectory}
\end{figure}

To understand the stationary behavior of DRABP, it is useful to characterize the long-time trajectories in the different phases.
\begin{itemize}

\item {\bf Passive-I}. A typical trajectory of DRABP in this phase, shown in Fig.~\ref{f:trajectory}(a), resembles that of an ordinary Brownian particle in a harmonic trap. This is because the randomization time-scale $D_R^{-1}$ of the orientation is much smaller than the relaxation time-scale $\mu^{-1}$ of the trap. Increasing $\gamma$ decreases the randomization time-scale to $(D_R+2\gamma)^{-1}$. Consequently, the description of DRABP at a time-scale larger than this randomization time-scale is given by an Ornstein-Uhlenbeck process with an effective diffusion constant $D_{\text{DR}}$.

\item {\bf Active-I}. Figure~\ref{f:trajectory}(b) shows a typical trajectory in this phase.  Except a very few detours to the interior region, the particle mostly stays near the boundary, where the net force on the particle is zero when its orientation vector $\hat{\bm{n}}$ is along $\bm{r}$. This is due to the fact that in this regime $\theta$ changes slowly as well as reversal events are very rare. 

\item{\bf Passive-II}. Figure~\ref{f:trajectory}(c) shows a typical trajectory in this regime. Unlike active-I, the large number of directional reversals makes the persistence length $\sim v_0/\gamma$ smaller than the diameter of the confining region $\sim v_0/\mu$. As a result, the particle is confined near the origin. However, unlike passive-I, since $D_R$ is small here, trajectory-segments between consecutive reversals are almost straight and pass through the central region, leading to a qualitatively different distribution. 

\item{\bf  Active-II}. As seen from Figure~\ref{f:trajectory}(d), since $D_R$ is small, in this regime also the particle passes through the central region almost in a straight line. However, unlike the passive-II, since the persistence length $\sim v_0/\gamma$ is larger than the diameter of the confining region $\sim v_0/\mu$, it goes all the way to the boundary and spends a considerable time there leading to a concentration of probabilities at the boundary as well as the center.

\end{itemize}
These four classes of different trajectories lead to four qualitatively different shapes of the stationary distribution, which we analyze in the following section.

\section{Stationary position distributions in the phases}\label{sec:ss_dist}
In this section we derive analytical expressions for the stationary distributions in the different phases. We also support this picture beyond the limiting cases using numerical simulations. Let us start with the most familiar passive phase (passive-I) where the stationary state is Boltzmann-like. 

\subsection{Passive-I phase: $D_R\gg\mu$}\label{s:pass1}
\begin{figure}[t]
\centering{\includegraphics[width= 0.9\hsize]{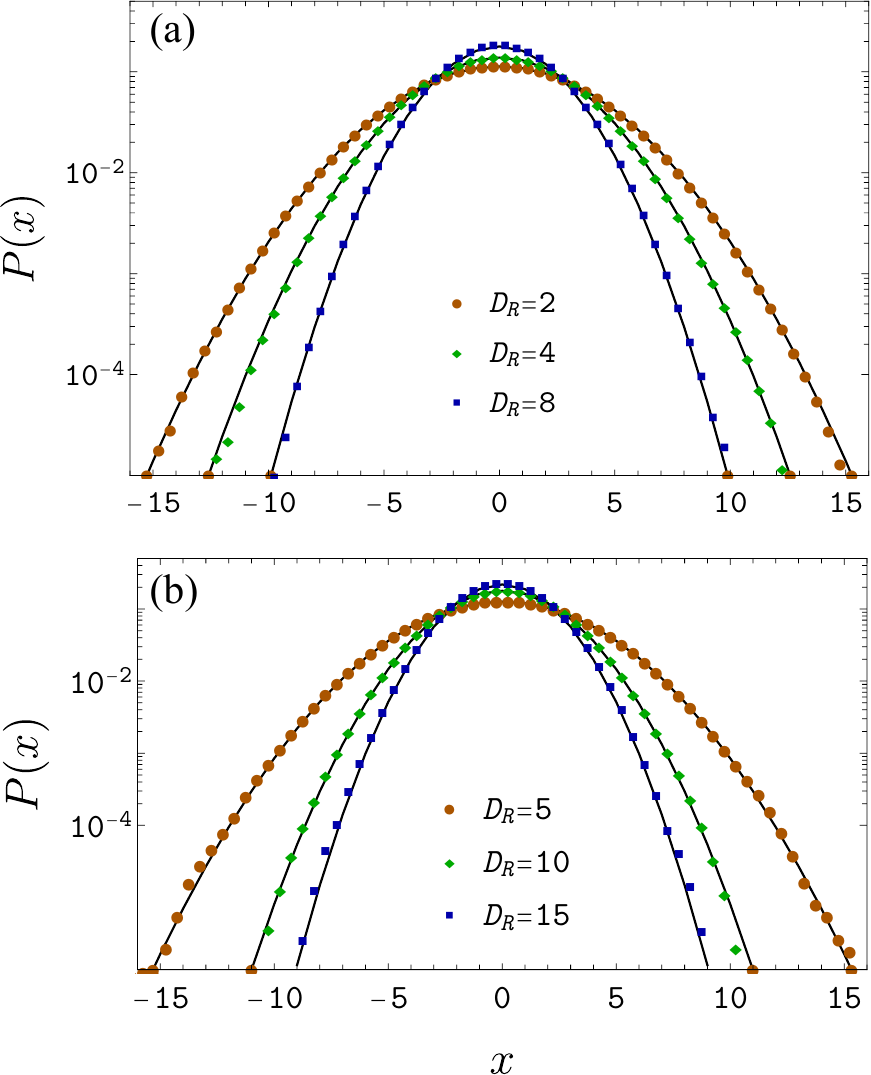}}
\caption{The stationary marginal distribution $P(x)$ in the passive-I phase for $\mu=0.01$ and $v_0=1$. The analytical predictions for $P(x)$ given by ~\eref{pxdgam-gaussian} and \eqref{abpgaussX} (solid black lines) are compared with numerical simulations (solid symbols) for $\gamma=1$ (a) and $\gamma=10^{-4}$ (b) respectively.}
\label{boltzmann1}
\end{figure}

In this case it is useful to rewrite Eq.~\eqref{langevin} as,
\begin{subequations}
\begin{align}
\dot{x}=-\mu x+\zeta_x(t),\\
\dot{y}=-\mu y+\zeta_y(t).
\end{align}
\label{eff-noise}
\end{subequations}
 The auto-correlation of the effective noises $\zeta_x(t)=v_0\sigma(t)\cos\theta(t)$ and $\zeta_y(t)=v_0\sigma(t)\sin\theta(t)$ become~\cite{drabp}
 \begin{align}
 \la \zeta_x(t)\zeta_x(t') \ra=\la\zeta_y(t)\zeta_y(t')\ra\to \frac{v_0^2}{2}e^{-(D_R+2\gamma)|t-t'|},
 \end{align}
 for $t$, $t'\gg D_R^{-1}$ and arbitrary $\gamma$, while the cross-correlation $\la \zeta_x(t)\zeta_y(t') \ra\to 0$. Now, for large $D_R$, we can evolve \eref{eff-noise} at a time step $D_R^{-1}\ll dt\ll \mu^{-1}$, where the effective noises emulate two independent white noises with  auto-correlations 
 \begin{align}
 \la \zeta_a(t)\zeta_b(t') \ra\to 2D_{\text{DR}}\,\delta_{a,b}\,\delta(t-t'),\text{ with } \{a,b\}\in\{x,y\}
 \end{align}
 and $D_{\text{DR}}=v_0^2/[2(D_R+2\gamma)]$.
 Thus, the Langevin equations (\ref{eff-noise}) reduce to an Ornstein-Uhlenbeck process, where the stationary state is given by the Boltzmann distribution,
 \begin{align}
P(x,y)&=\frac{\mu }{2\pi D_{\text{DR}}}\,\exp\left[-\frac{\mu (x^2+y^2)}{2D_{\text{DR}}}\right].
\label{gaus_drabpXY}
\end{align} 
This is the passive-I phase [see Fig.~\ref{2dplot}(a)] as announced in Secs.~\ref{sec:model} and \ref{sec:trajectory}. The corresponding marginal distribution is evidently also a Gaussian,
\begin{align}
P(x)&=\sqrt{\frac{\mu }{2\pi D_{\text{DR}}}}\,\exp\left(-\frac{\mu x^2}{2D_{\text{DR}}}\right).
\label{pxdgam-gaussian}
\end{align} 
which is compared with the numerical simulations in
Figure~\ref{boltzmann1}(a).

Note that, the variance of the above distribution agrees with the expression obtained from the exact \eref{ss_var} by taking the limit $v_0\to\infty$, $D_R\to\infty$ keeping $v_0^2/D_R$ constant for arbitrary $\gamma$. Moreover, the exact kurtosis given by \eref{kurt} tends to zero in this limit, consistent with the Gaussian form of the above distribution [see \eref{mo:pass1}].

  In the limit $\gamma\to 0$, the above distribution reduces to that of an ABP in a harmonic potential in the passive phase~\cite{abp2018}, 
  \begin{align}
P(x,y)&=\frac{\mu }{2\pi D_{\text{AB}}}\,\exp\left[-\frac{\mu (x^2+y^2)}{2D_{\text{AB}}}\right],
\label{abpgaussXY}
\end{align}
 where $D_{\text{AB}}=v_0^2/(2D_R)$. The corresponding marginal distribution is also obviously a Gaussian,
 \begin{align}
P(x)&=\sqrt{\frac{\mu }{2\pi D_{\text{AB}}}}\,\exp\left(-\frac{\mu x^2}{2D_{\text{AB}}}\right),
\label{abpgaussX}
\end{align} 
which we compare with numerical simulations in Fig.~\ref{boltzmann1}(b).

Next we discuss the most commonly seen active phase where the stationary probability density is concentrated near the boundary.

\subsection{Active-I phase: $\gamma\ll D_R\ll\mu$}\label{sec:gam0}
In the limit $\gamma/D_R\to 0$, the Fokker-Planck equation \eqref{eq:fp} becomes,
\begin{align}
\frac{\partial P_{\sigma}}{\partial t}=&-\left[\frac{\partial}{\partial x}(-\mu x+v_0\sigma\cos\theta)+\frac{\partial}{\partial y}(-\mu y+v_0\sigma\sin\theta)\right]P_{\sigma}\nonumber\\ &+D_R\frac{\partial^2P_{\sigma}}{\partial \theta^2},
\label{fp:gam0}
\end{align}
where $P_{\pm}$ represents two non-interacting ABPs with constant velocities $v_0$ and $-v_0$ respectively. Since the stationary state of an ABP does not depend on the sign of the velocities (alternatively, the initial orientation), in this limit we get the same stationary distributions as that of an ABP in a harmonic potential~\cite{abp2018,abp2020}.

\begin{figure}
\centering{
\includegraphics[width=0.9\hsize]{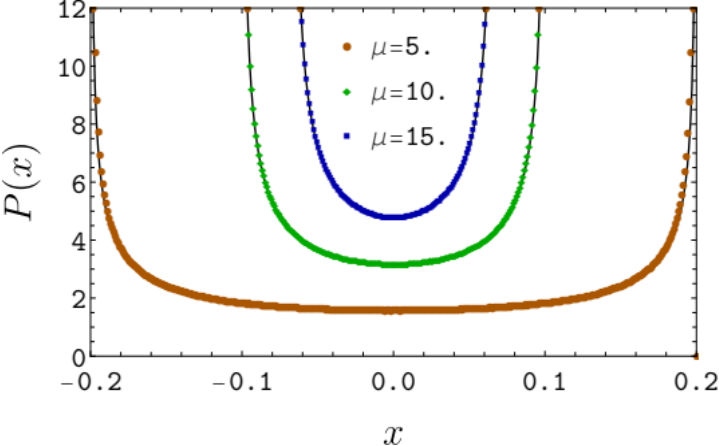}}
\caption{The stationary marginal distribution $P(x)$ in the active-I phase. The symbols are obtained from  numerical simulations for $D_R=0.05$ and $\gamma=10^{-4}$  and solid black lines are from the analytical prediction \eref{eq:gsmall1}.}
\label{abp_plane}
\end{figure}

 Therefore, for $\gamma\ll D_R\ll\mu$, the probability density is concentrated along a ring $x^2+y^2=r_0^2$ thereby indicating that the particle is in an active phase.~\footnote[3]{Note that, the Fokker-Planck equation \eqref{fp:gam0} still holds for $\gamma\ll\mu\ll D_R$. The corresponding passive phase (passive-I) is characterized by a Boltzmann distribution as given in \eref{abpgaussXY}.} In other words,  the position distribution takes the limiting form 
\begin{align}
P(x,y)&=\frac{1}{2\pi r_0}\delta\left(\sqrt{x^2+y^2}-r_0\right).
\label{eq:gsmall1}
\end{align}
We refer to this ABP-like active phase of DRABP as active-I to distinguish it from a novel active phase obtained in Sec.~\ref{sec:dr0}, emerging from the direction reversal. The marginal distribution is obtained by integrating  Eq.~\eqref{eq:gsmall1} over $y$ as,
\begin{align}
P(x)&=\frac{1}{\pi\sqrt{r_0^2-x^2}}\,\Theta(r_0-|x|),
\label{active:marg}
\end{align}
where the $\Theta(z)$ is the Heaviside-theta function.
We compare this theoretical prediction with  numerical simulations in Fig.~\ref{abp_plane} and find an excellent agreement. Interestingly, this active-I phase even extends to $\gamma/D_R= O(1)$, where the shape of the distribution remains qualitatively same (weighted near the boundary), as discussed later in Sec.~\ref{numericalss}. The variance and the kurtosis corresponding to \eref{active:marg} are $r_0^2/2$ and $-3/2$ respectively, which are consistent with the direct calculation of the same; see \eref{mo:act1}.

 Finally, we discuss two novel phases, where the directional reversal leads to  a diverging central peak along with the active and passive like features.

\subsection{The novel active and passive phases: $D_R\ll(\mu,\gamma)$} \label{sec:dr0}
The directional reversal leads to two new phases which are best seen in the $D_R/\mu\to 0$ limit. It is useful to divide both sides of \eref{eq:fp} by $\mu$, which gives,
\begin{align}
\frac{\partial P_{\sigma}}{\partial (\mu t)}=&-\left[\frac{\partial}{\partial x}\left(- x+r_0\sigma\cos\theta\right)+\frac{\partial}{\partial y}\left(- y+r_0\sigma\sin\theta\right)\right]P_{\sigma}\nonumber\\
 &-\nu\, \left[P_{\sigma}- P_{-\sigma}\right]+\frac{D_R}{\mu}\frac{\partial^2P_{\sigma}}{\partial \theta^2},
\label{fpdr0}
\end{align}
where $\nu=\gamma/\mu$.
In the limit $D_R/\mu\to 0$ while keeping $r_0$ and $\nu$ finite, $\theta$ evolves very slowly. As a first approximation, $\theta$ can be kept fixed. This is equivalent to neglecting the second order derivative with respect to $\theta$ in \eref{fpdr0}, resulting in the Fokker-Planck equation for the conditional distribution $P_{\sigma}(x,y,t|\theta)$ for a given $\theta$. 

 Now, for a fixed $\theta$, it is convenient to make a rotation of the coordinate system
\bea
\begin{bmatrix}
x_{\parallel}\\x_{\perp}\end{bmatrix}=\begin{bmatrix}
\cos\theta & \sin\theta\\
-\sin\theta & \cos\theta
\end{bmatrix} \begin{bmatrix}
x\\y\end{bmatrix},
\eea
where $x_{\parallel}$ and $x_{\perp}$ are respectively the axes parallel and perpendicular to the $\theta$-direction. In the $(x_{\parallel},x_{\perp})$ coordinates, the Fokker-Planck equation for $P_{\sigma}(x_{\parallel},x_{\perp},t|\theta)$ becomes,
\begin{align}
\frac{\partial P_{\sigma}}{\partial (\mu t)}=&-\frac{\partial}{\partial x_{\parallel}}\left[\left(- x_{\parallel}+r_0\sigma\right)P_{\sigma}\right]-\nu\left[ P_{\sigma}- P_{-\sigma}\right]-\frac{\partial}{\partial x_{\perp}}\left[-x_{\perp}P_{\sigma}\right].
\end{align}
It is evident from the above equation that the dynamics of $x_{\parallel}$ is nothing but that of a one-dimensional RTP along $\theta$ in a harmonic potential. On the other hand, $x_{\perp}$ independently undergoes a deterministic overdamped motion in a harmonic potential, resulting in $x_{\perp}\to 0$ as $t\to\infty$. Therefore, the steady state position distribution $P(x_{\parallel},x_{\perp}|\theta)=\sum_{\sigma=\pm 1}P_{\sigma}(x_{\parallel},x_{\perp},t\to\infty|\theta)$ can be obtained using the steady state result of 1D RTP in a harmonic trap~\cite{1drtptrap},
\bea
P(x_{\parallel},x_{\perp}|\theta)=\frac{\delta(x_{\perp}) \,2^{1-2\nu}}{r_0B(\nu,\nu)}\left[1-\left(\frac{ x_{\parallel}}{r_0}\right)^2\right]^{\nu-1}\Theta(r_0- |x_{\parallel}|),
\label{eq:1drtp}
\eea
where $B(\nu,\nu)$ is the beta function.

\begin{figure*}[htp]
\centering
\hspace{-0.84 cm}\includegraphics[scale=0.6]{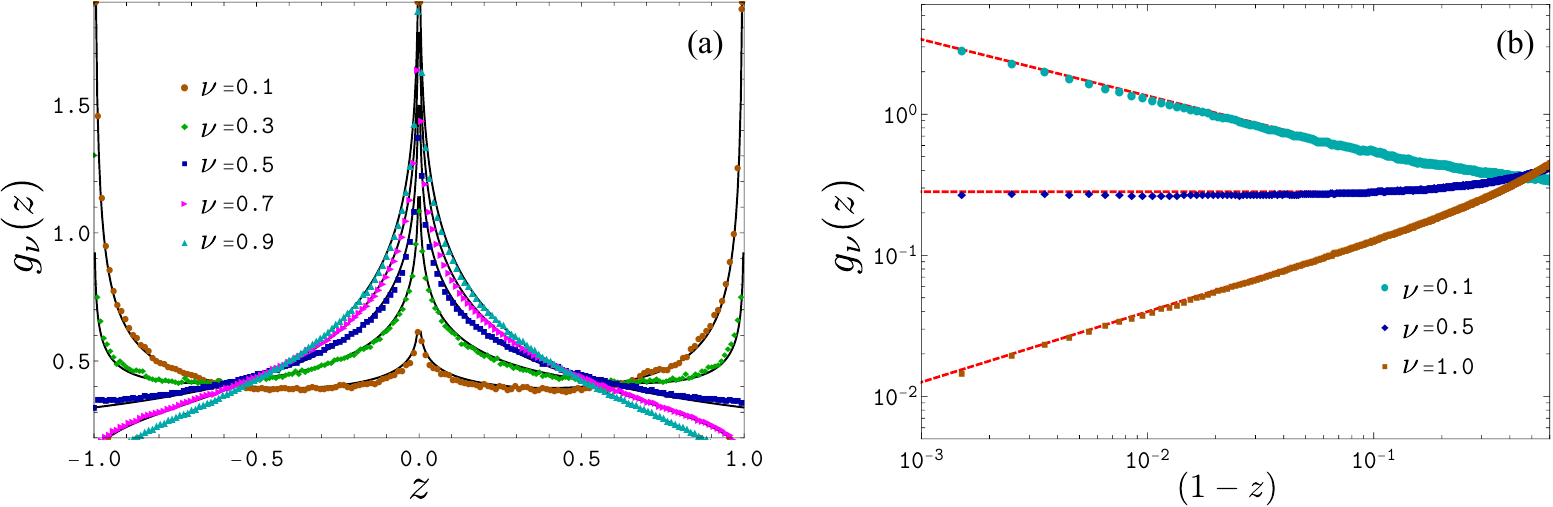}
\caption{(a) Comparison of theoretical stationary state marginal distribution $g_{\nu}(z)$ given by \eqref{marg_sc} as a function of the scaled variable $z=\mu x/v_0$ in the $D_R\to 0$ limit (solid black lines) with numerical simulations (symbols) for $D_R=10^{-4}$, and $\mu=v_0=1$, for different values of $\nu=\gamma/\mu$. This plot shows the transition from the Active-II to Passive-II at $\gamma=\mu/2$ as predicted by \eref{scalingtrans}. (b) Comparison of \eref{scalingtrans} (red dashed lines) with the simulations (symbols) for the same set of remaining parameters as in (a), which highlights the different tail behavior in the two phases. }
\label{fig:dr0}
\end{figure*}

Subsequently, in terms of the original coordinates $(x,y)$, the position distribution becomes,
\begin{align}
P(x,y|\theta)&=\frac{2^{1-2\nu}}{r_0 B(\nu,\nu)}\left[ 1- \frac{x^2+y^2}{r_0^2} \right]^{\nu-1} \cr &\times \delta(- x \sin\theta + y \cos\theta)\,\Theta\left(r_0- \sqrt{x^2 + y^2}\right).
\label{pxydr0}
\end{align}

The dynamics of $\theta$ is independent of that of $(x,\,y)$, whose distribution evolves by the diffusion equation, leading to the uniform steady state for $\theta\in[0,2\pi]$ for $t\gg D_R^{-1}$. Averaging \eref{pxydr0} with respect to the steady state distribution $\theta$, we get the scaling form for the distribution,
\bea 
P(x,y)=\int_{0}^{2\pi}\frac{d\theta}{2\pi} P(x,y|\theta) =\frac{1}{ r_0^2}f_{\nu}\left( \frac x{r_0}, \frac y{r_0}\right),
\label{pxyscaled}
\eea
with the scaling function,
\begin{align}
f_{\nu}(z_1,z_2)&=\frac{2^{1-2\nu}}{\pi B(\nu,\nu)}\frac{(1-z_1^2-z_2^2)^{\nu-1}}{\sqrt{z_1^2+z_2^2}}\Theta(1-z_1^2-z_2^2).
\label{eq:2dsc}
\end{align}
Plots of the scaling distribution $f_{\nu}(z_1,z_2)$ are shown in Fig.~\ref{2dplot}(c) and (d) for $\nu>1$ and $\nu<1$ respectively. For $\nu<1$, the distribution looks like a Mexican hat with algebraic divergences both at the origin and at the boundary $z_1^2+z_2^2=1$. On the other hand for $\nu>1$, the distribution goes to zero at the boundaries, while it still retains the algebraic divergence at the origin.

The marginal distribution can be obtained by integrating Eq.~\eqref{pxyscaled} over one of the coordinates. Integrating over $y$ yields the scaling form,
\bea
P(x)=\frac{1}{r_0}g_{\nu}\left(\frac{x}{r_0}\right).
\label{marg:dr0}
\eea
The corresponding scaling function is given by,
\begin{align}
g_{\nu}(z)=&\frac{1}{\pi}(1-z^2)^{\nu-\frac 12} {}_2F_1\left(\frac12,\nu,\nu+\frac12,1-z^{2}\right)\Theta(1-z^2),
\label{marg_sc}
\end{align}
where $_2F_1(a,b,c,y)$ denote the Hypergeometric function. In fact, one can generalize the above result for the marginal distribution to all higher dimensions [see \eref{e:ddim}], as shown in Appendix~\ref{sec:ddim}. Incidentally, the radial distribution is independent of dimensionality and is given by \eref{rad_dist}.

The moments of the above distribution can be computed by using the series representation of the hypergeometric function. The variance $\la x^2\ra=r_0^2/[2(2\nu+1)]$ and the kurtosis $\kappa=3(2\nu-3)/[2(2\nu+3)]$ obtained from \eref{marg_sc} agree with the direct calculations of the same (see \eref{mo:pass2}).

\Fref{fig:dr0}(a) shows a very good agreement between \eref{marg_sc} and numerical simulations for small values of $D_R$. As seen in Fig.~\ref{fig:dr0}(a), the shape of the distribution near the boundaries shows three qualitatively different behaviors. Indeed, it follows from Eq.~\eqref{marg_sc} that the behavior of the tails near  $z=\pm 1$ undergoes a transition as a function of $\nu$,
\bea
g_{\nu}(z)\simeq\frac 1{\pi}\times\begin{cases}
	\big[2(1-|z|)\big]^{-(1/2-\nu)}\quad & 0<\nu<1/2,\\      
 1 \quad &\nu=1/2,\\     
      \big[2(1-|z|)\big]^{\nu-1/2}\quad  &\nu>1/2.\\
          \end{cases}
\label{scalingtrans}
\eea
It is evident from the above equation that at the boundaries $z=\pm 1$, the marginal distribution diverges for $\nu<1/2$, while it vanishes for $\nu>1/2$. We compare this theoretical prediction with numerical simulations in Fig.~\ref{fig:dr0}(b) and find excellent agreement.

\begin{figure}
\centering{
\includegraphics[width=0.9\hsize]{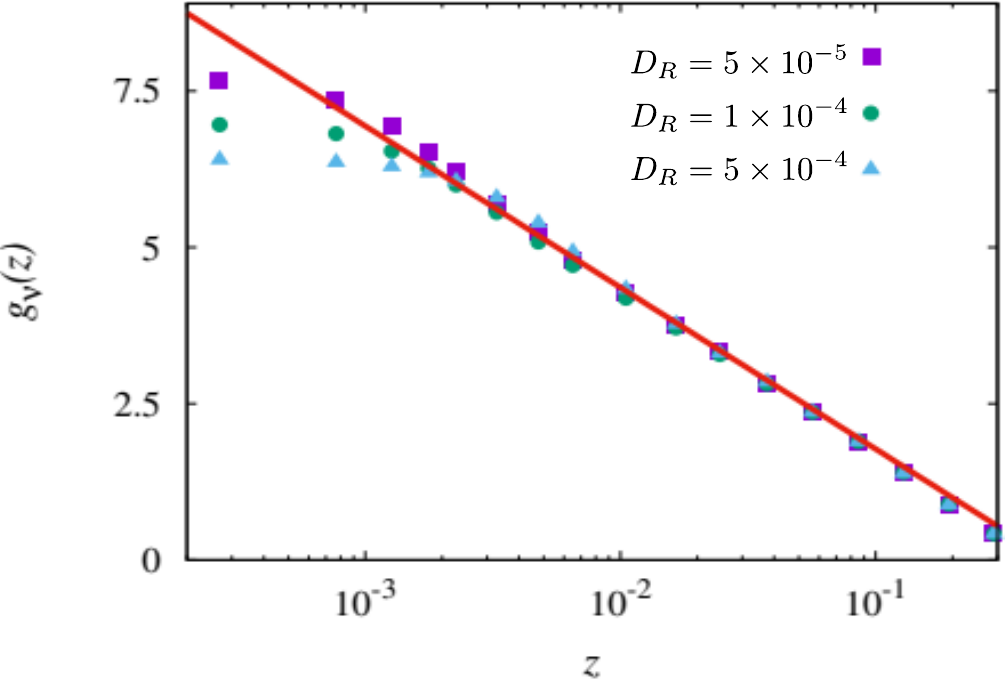}}
\caption{The central logarithmic divergence for the marginal stationary state distribution in the passive-II phase. The symbols denote the scaled distribution obtained using numerical simulations for $\gamma=10$, $\mu=v_0=1$, while the solid red line indicates the analytical prediction in \eref{centredr0}. We see progressively better agreement for smaller values of $D_R$.} 
\label{f:centralsc}
\end{figure}

One distinctive feature of the scaling function in \eref{marg_sc} is that, for all values $\nu$, it has a logarithmic divergence at the center,
\begin{align}
g_{\nu}(z)&=-\frac{\Gamma \left(\nu +\frac{1}{2}\right) }{\pi ^{3/2} \Gamma (\nu )}\,\left[\log \left(\frac{z^2}{4}\right)+E+\psi(\nu )\right]+O(z^2),
\label{centredr0}
\end{align}
where  $\Gamma(\nu)$ is the gamma function, $E=0.5772\dots$ is the Euler-Mascheroni constant, and $\psi(\nu )=\Gamma'(\nu)/\Gamma(\nu)$ is the digamma function. We illustrate the above small $z$ behavior of $g_{\nu}(z)$ and compare it with numerical simulation in Fig.~\ref{f:centralsc}. As expected, we find progressively better agreement for smaller values of $D_R/\mu$ for a fixed value of $\nu$. 

The divergence of $f_{\nu}(z_1,z_2)$ in \eref{eq:2dsc} at the boundary $z_1^2+z_2^2=1$ for $\nu<1$ is a signature of activity, implying the accumulation of particles near the boundary. However, this phase [see Fig.~\ref{2dplot} (d)] is different from the active-I phase discussed earlier [see Fig.~\ref{2dplot} (b)] marked by the presence of an additional central diverging peak. We refer to this phase as active-II. On the other hand, the distribution has only a central peak for $\nu>1$, which is characteristic of the passive phase. However, the diverging nature of the central peak distinguishes this phase [see Fig.~\ref{2dplot} (c)] from the usual passive-I phase [see Fig.~\ref{2dplot} (a)] discussed earlier. We refer to this phase as passive-II. Note that, the transition from active-II to passive-II occurs at $\nu=1/2$ for the marginal distribution [see \eref{scalingtrans}], in contrast to $\nu=1$ for the two-dimensional joint distribution. Moreover, in higher dimensions $d\geq 3$, the marginal distribution $g_{d,\nu}(z)\propto(1-z^2)^{\nu+(d-3)/2}$ [see \eref{e:ddim}], does not show any boundary accumulation. Therefore, in $d\geq3$, the signature of the active phase is present only in the full $d$-dimensional distribution and the radial distribution \eref{rad_dist}.

To further highlight the novel features of the passive-II phase in $d=2$, we analyze the position distribution \eqref{eq:2dsc} in the typical diffusive scaling limit of RTP,  $\gamma\to\infty$ and  $v_0\to\infty$ while keeping $v_0^2/\gamma=2D_{\text{RT}}$ fixed [also see Appendix~\ref{sec:moment}]. This is equivalent to taking the limit $\nu\to\infty$ and $z_1$, $z_2\to 0$, keeping $z_1\sqrt{\nu}$ and $z_2\sqrt{\nu}$ finite. This leads to the scaling form,
\begin{align}
f_{\nu}(z_1,z_2)=\nu\, h\left(z_1\sqrt{\nu},z_2\sqrt{\nu}\right),
\end{align} 
and consequently, $P(x,y)$ has the scaling form,
\begin{align}
P(x,y)=\frac{\mu}{2D_{\text{RT}}}h\left(x\sqrt{\frac{\mu}{2D_{\text{RT}}}},y\sqrt{\frac{\mu}{2D_{\text{RT}}}}\right).
\label{eq:2dscaled}
\end{align}
The corresponding scaling function is given by,
\begin{align}
h(w_1,w_2)=\frac{1}{\pi^{3/2}}\frac{\exp(-w_1^2-w_2^2)}{\sqrt{w_1^2+w_2^2}},\,~~\{w_1,w_2\}\in(-\infty,\infty).
\label{eq:sc2d}
\end{align}
The normalization $\int_{-\infty}^{\infty}dw_1\int_{-\infty}^{\infty}dw_2\, h(w_1,w_2)=1$ is easily checked.

 While $P(x,y)$ has the Boltzmann tail $\propto \exp\left[- V(x,y)/D_\text{RT}\right]$ with the potential $V(x,y)=\mu (x^2+y^2)/2$, it has a novel algebraic divergence at the origin, unlike the passive-I case. By integrating \eqref{eq:2dscaled} over $y$, we get the marginal distribution, 
 \bea
P(x)=\sqrt{\frac{\mu}{2D_{\text{RT}}}}\,q\left(x\sqrt{\frac{\mu}{2D_{\text{RT}}}}\right),
\eea
where the scaling function $q(w)=\int_{-\infty}^\infty h(w,w')\,dw'$ is given by,
\bea
q(w)=\frac{1}{\pi^{3/2}}\,K_0\left(\frac{w^2}{2}\right)\exp\left(-\frac{w^2}{2}\right).
\label{origin_scale}
\eea
Here $K_0(z)$ is the zeroth order modified Bessel function of second kind and the normalization $\int_{-\infty}^\infty q(w)\,dw=1$ is easily checked. 

\begin{figure}
\centering{
\includegraphics[width= 0.9\hsize]{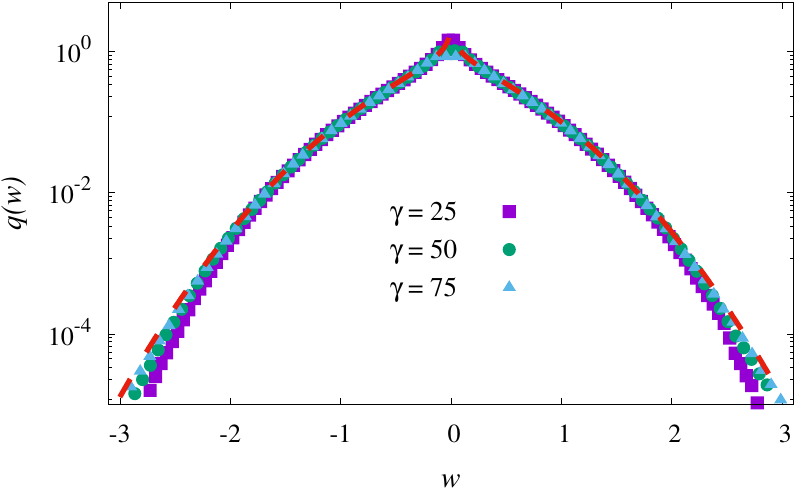}}
\caption{The stationary marginal distribution in the passive-II phase in the scaling limit $\gamma\to\infty$, $v_0\to\infty$ keeping $D_{\text{RT}}=v_0^2/(2\gamma)$ finite, as a function of the scaled variable $w=x\sqrt{\mu}/\sqrt{2D_{\text{RT}}}$. The symbols denote the data obtained from numerical simulations, while the dashed red line corresponds to the scaling function given by \eref{origin_scale}. As expected, departure from the predicted scaling behavior occurs for larger values of $w$ and increases with lower values of $\gamma$. We have used $\mu=1$, $v_0=10$ and $D_R=10^{-4}$.}
\label{f:logscaling}
\end{figure}

 The asymptotic behavior $K_0(w^2/2)\sim\exp(-w^2/2)$ as $w\to\infty$, leads to the Boltzmann distribution at the tails as expected. However, the small-$w$ behavior $K_0(w^2/2)=-[\log(w^2/4)+E]+O(w^4)$, leads to a logarithmic divergence,
\begin{equation}
q(w)=-[\log(w^2/4)+E]+O(w^2),
\label{origin_scale}
\end{equation}
near the origin. This is in agreement with \eref{centredr0} for large $\nu$ and taking  $w=z\sqrt{\nu}$ as the scaling variable.

\subsection{Crossover from active-I to active-II}\label{numericalss}
As discussed in Sec.~\ref{s:pass1}, the scenario where $\mu^{-1}$ is the largest among the three time-scales yields the passive-I phase. On the other hand, the complementary scenario where $\mu^{-1}$ is the smallest time-scale, can lead to both active-I and active-II phases, as discussed earlier in Secs.~\ref{sec:gam0} and \ref{sec:dr0} respectively. It arises from the two limits of the Fokker-Planck equation \eqref{fpdr0} in the rescaled time ($\mu t$): it leads to the active-I phase for $\gamma/D_R\to 0$, while for  $D_R/\gamma\to 0$, it gives the active-II phase. To understand the crossover from the active-I to the active-II, as $\gamma/D_R$ is varied, we take recourse to numerical simulations and study the phase diagram on the $(\gamma,\,D_R)$ planes for fixed values of $\mu$. 

We scan the $(\gamma,\,D_R)$ plane  for a range of values of $\gamma$ and $D_R$ at an interval of $\Delta\gamma=\Delta D_R=0.05$, and obtain the marginal stationary state distribution from simulation at these points.
To distinguish between the different phases, we numerically detect the existence of the peaks near the origin and the boundaries.
To detect if there is a peak away from the origin, we check, whether, for some $\epsilon\ll r_0$, the first order finite difference $\rho(x_0)-\rho(x_0-\epsilon)$ is positive for some $x_0\in(0,r_0)$ --- suggesting  $\rho(x)$ increases with $x$, and thus, there is an accumulation away from the origin. This is the signature of active phase. Now, to differentiate between active-I and active-II, we further check the existence of an additional peak at the origin.
This central peak is detected by monitoring the sign of the second order finite difference $\rho(-\epsilon)+\rho(\epsilon)-2\rho(0)$ , where the negative sign corresponds to a maximum  at the origin. 

\begin{figure*}[htp]
\centering
\hspace{-0.5 cm}\includegraphics[width=0.9\hsize]{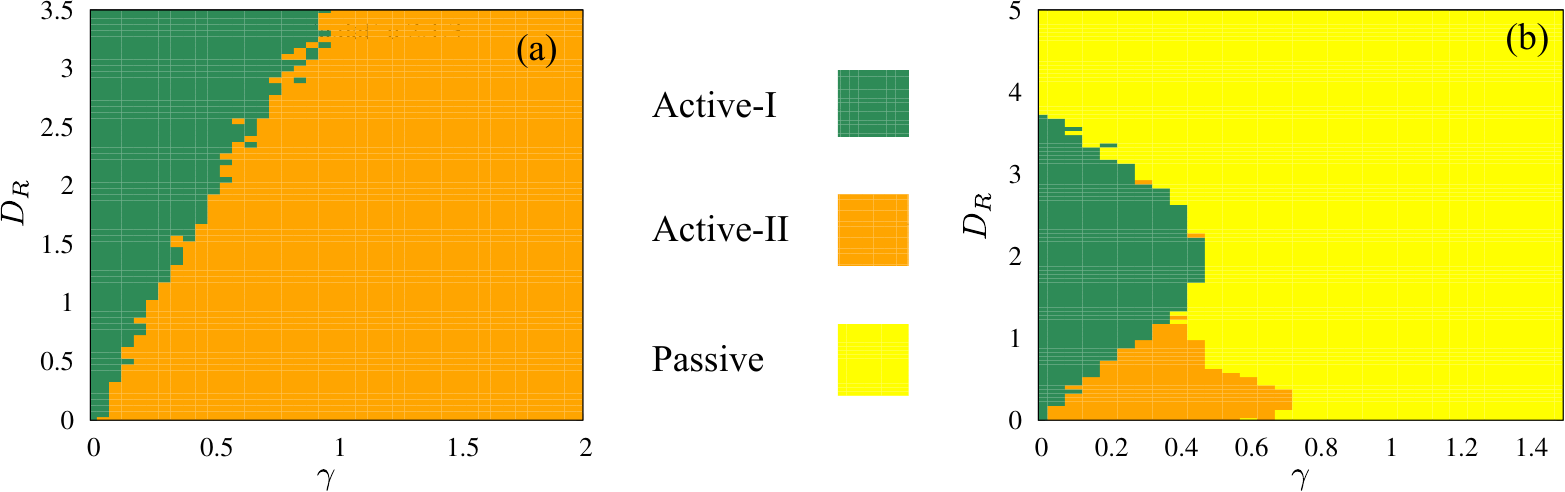}
\caption{The different phases in the $(\gamma,\,D_R)$ plane for $\mu=10$ (a) and $\mu=2$ (b). (a) shows the phases active-I (green region) and active-II (orange region) detected numerically using the procedure described in Sec.~\ref{numericalss}. This phase plane is schematically represented in Fig.~\ref{phasediagram} (c). The passive phases appear for larger values of $\gamma/\mu$ and $D_R/\mu$, which is visible (yellow region) in (b) for $\mu=2$.}
\label{f:u}
\end{figure*}

We use this method with $\epsilon=0.02$ to obtain the phase diagram shown in Fig.~\ref{f:u}, which illustrates the transition (a) from active-I to active-II for $\mu=10$ and (b) between active and passive phases for $\mu=2$. We make two observations from our numerically obtained phase diagram. First, the boundary between the active-I and active-II phases is almost linear and passes through the origin. Secondly, the cross-sectional area corresponding to the combined active regions shrinks with increasing $\mu^{-1}$, thus implying a funnel-like surface. However, 
the shape of the boundary between the active and passive phases suggest that the surface of the funnel may have more complex structure than the simple schematic representation shown in Fig.~\ref{phasediagram}(a).

\section{Conclusion}\label{sec:concl}
In conclusion, in this paper we have studied the stationary state of an active Brownian particle with intermittent directional reversals, in the presence of a harmonic trap.
The interplay of the rotational diffusion constant $D_R$, the reversal rate $\gamma$, and the trap strength $\mu$  leads to a complex phase diagram. We classify the different phases by obtaining the exact analytical expressions for the position distribution in the limiting scenarios.

  We find that for $D_R\gg\mu$, the system always relaxes to a Boltzmann-like distribution, for any $\gamma$, which we refer to as the passive-I phase. On the other hand, for $D_R\ll\mu$, depending on the strength of $\gamma$ relative to the two other parameters, we get three different phases: For $\gamma\ll D_R$, an ABP-like active phase emerges where the particle is most likely to be found near the circular boundary of radius $v_0/\mu$. This is referred to as the active-I phase. For $D_R\ll\gamma<\mu$, a new active phase (active-II) emerges which crosses over to a new passive phase (passive-II) for $\gamma>\mu$. Both these new phases are characterized by the presence of a non-Gaussian, diverging central peak of the position distribution. However, the active-II is distinguished from the passive-II phase by the presence of the typical accumulation of probability density near the circular boundary. The exact position distribution obtained analytically in the $D_R\to 0$ limit yields the exact transition line in the $(\gamma,\,\mu^{-1})$ plane. 
  Finally, we complete the phase diagram by studying the characteristic shape of the position distribution numerically, beyond the analytically accessible limits.

This work gives rise to several interesting open questions. It would be interesting to look at the first-passage properties of DRABP in presence of a harmonic trap, as it exhibits novel persistence behavior in the absence of any confinement.
  Another natural question is how the phase diagram changes for a generic  force of the form $F(\vec{r})=-\hat{r}\,r^p$. Incidentally, the harmonic potential corresponds to the case $p=1.$ Since DRABP is a minimal model for a certain class of bacterial motion, it would be intriguing to find out whether the different phases predicted here can be observed in experiments with bacteria like {\it Myxococcus xanthus, Pseudomonas
citronellolis} etc., or light activated Janus colloids~\cite{Verman2020} in harmonic trap.

\appendix
\section*{Appendix}
\section{Variance and kurtosis of the position distribution}\label{sec:moment}

Here we calculate the variance and kurtosis of the DRABP in a harmonic trap exactly. We use these to find the limiting expressions in the four different phases.
Starting from the origin, the solution of the Langevin equation \eqref{langevin}, for a given realization of $\{\sigma(s),\theta(s);0<s<t\}$, gives the location $\{x(t),y(t)\}$ of the DRABP as,
\begin{subequations}
\begin{align}
x(t)&=v_0\int_{0}^{t} ds\,e^{-\mu(t-s)}\sigma (s) \cos\theta (s), \\
y(t)&=v_0\int_{0}^{t} ds\,e^{-\mu(t-s)}\sigma (s) \sin\theta (s). 
\end{align}\label{int:langevin}
\end{subequations}

 We consider the initial condition $\sigma(0)=\pm 1$ with equal probability $1/2$, which ensures that all the odd moments of the position vanish at all times. To calculate the first nontrivial moment, the variance, we need the two point correlations of the noises~\cite{drabp},
\begin{subequations} 
 \begin{align}
\la \sigma(s)\sigma(s')\ra \la \cos\theta(s)&\cos\theta(s')\ra = \frac 12\left[e^{-(D_R+2\gamma)|s-s'|}\right.\cr &+ \left.e^{-2\gamma|s-s'|-D_R(s+s'+2\,\text{min}[s,s'])}\right], \\
\la \sigma(s)\sigma(s')\ra \la \sin\theta(s)&\sin\theta(s')\ra= \frac 12\left[e^{-(D_R+2\gamma)|s-s'|}\right.\cr &+ \left.e^{-2\gamma|s-s'|-D_R(s+s'+2\,\text{min}[s,s'])}\right],
 \end{align}\label{noise_c}
\end{subequations}
for the initial condition $\theta(0)=0$.
Using~\eref{int:langevin} and \eqref{noise_c} we obtain,

\begin{align}
\langle x^2(t)\rangle &=\frac{v_0^2}{2 \mu  \left(2 \gamma +D_R+\mu \right)}-\frac{ v_0^2\left(D_R-\mu \right)\,e^{-2 \mu  t}}{\mu  \left(2 D_R-\mu \right) \left(2 \gamma +D_R-\mu \right)}\cr 
&+\frac{2v_0^2 \left(2 \gamma -D_R+\mu \right)\, e^{- \left(2 \gamma +D_R+\mu \right)t}}{\left(2 \gamma +D_R-\mu \right) \left(2 \gamma +D_R+\mu \right) \left(2 \gamma -3 D_R+\mu \right)}\cr
 &-\frac{v_0^2\,e^{-4  D_Rt}}{2 \left(2 D_R-\mu \right) \left(2 \gamma -3 D_R+\mu \right)}, \label{varx}
\end{align}
and,
\begin{align}
\langle y^2(t)\rangle &=\frac{v_0^2}{2 \mu  \left(2 \gamma +D_R+\mu \right)}-\frac{v_0^2D_R\, e^{-2 \mu  t}}{\mu  \left(2 D_R-\mu \right) \left(2 \gamma +D_R-\mu \right)}\cr 
&-\frac{4v_0^2 D_R\, e^{- \left(2 \gamma +D_R+\mu \right)t}}{\left(2 \gamma +D_R-\mu \right) \left(2 \gamma +D_R+\mu \right) \left(2 \gamma -3 D_R+\mu \right)}\cr
 &+\frac{v_0^2\,e^{-4  D_Rt}}{2\left(2 D_R- \mu \right) \left(2 \gamma -3 D_R+\mu \right)}.
\label{vary}
\end{align}
%
In the limit $t\to\infty$, both $\langle x^2(t)\rangle$ and $\langle y^2(t)\rangle$ relax to the same stationary value,
\begin{align}
\la x^2(t\to\infty)\ra=\la y^2(t\to\infty)\ra=\frac{v_0^2}{2 \mu  \left(2 \gamma +D_R+\mu \right)}+O(e^{-\lambda t}),
\label{ss_var}
\end{align}
where $\lambda=\text{min}(2\mu,\,4D_R,\, 2\gamma+D_R+\mu )$ gives the leading order time-scale of relaxation to the stationary value.

For a Gaussian distribution, all the higher order cumulants $\la x^n\ra_c$ with $n>2$ are zero. The widely used measure to identify non-Gaussianity is the fourth cumulant  $\la x^4\ra_c$, which is also known as kurtosis. It is often expressed in the dimensionless form,
\begin{align}
\kappa(t)&=\frac{\la x^4(t)\ra-3\la x^2(t)\ra^2}{\la x^2(t)\ra^2}.
\end{align}
Note that, however, vanishing kurtosis is not a sufficient condition for Gaussianity.
Clearly, to compute the kurtosis, we need the fourth moment of the distribution, which can be calculated using \eref{int:langevin} and the four-point correlations of the noises. Using the propagators [eqn.~(2) and (4) of the Supplementary Material of Ref.~\cite{drabp}] these four-point correlations can be calculated in a straightforward manner. For $t_1<t_2<t_3<t_4$, 
\begin{align}
\la\sigma(t_1)&\sigma(t_2)\sigma(t_3)\sigma(t_4) \ra=e^{-2 \gamma\,  (t_4-t_3)}e^{-2 \gamma\,  (t_2-t_1)},\label{sig4pt}
\end{align}
and for the $\theta$-process,
\begin{align}
\la\cos\theta(t_1)&\cos\theta(t_2)\cos\theta(t_3)\cos\theta(t_4)\ra =\frac{1}{8} e^{-D_R\left(7 t_1+5 t_2+3 t_3+t_4\right) }\cr
&\times\left(e^{12D_R t_1 }+e^{8  D_R\left(t_1+t_2\right)}+e^{4  D_R\left(t_1+2 t_2\right)}\right.\cr
& \left.+2 e^{4  D_R\left(t_1+t_2+t_3\right)} +2 e^{4  D_R\left(2 t_1+t_2+t_3\right)}+1\right). \label{cos4pt}
\end{align}
The full time-dependent fourth moment has a fairly large expression which upon taking the $t\to\infty$ limit yields,
\begin{align}
\la x^4(t\to\infty)\ra&=\frac{3 \left(4 D_R+3 \mu \right)}{8 \mu ^2 \left(2 D_R+\mu \right) \left(2 \gamma +D_R+\mu \right) \left(2 \gamma +D_R+3 \mu \right)}.
\end{align}
 The stationary state value of the kurtosis can be readily obtained using the second and fourth moments derived above, and comes out to be,
\begin{align}
\kappa(t\to\infty)&=\frac{3 \mu  \left(2 \gamma -7 D_R-3 \mu \right)}{2 \left(2 D_R+\mu \right) \left(2 \gamma +D_R+3 \mu \right)}.~~~~
\label{kurt}
\end{align}

The limiting expressions of variance and kurtosis in the different phases can be easily obtained from \eref{ss_var} and \eqref{kurt} respectively:
\begin{itemize}
\item In the limit $v_0\to\infty$, $D_R\to\infty$ with arbitrary $\gamma$ and $\mu\ll D_R$, keeping $v_0^2/(D_R+2\gamma)=2D_{\text{DR}}$ constant (passive-I phase), we get
\begin{equation}
\la x^2\ra=\frac{D_{\text{DR}}}{\mu}\quad\text{and}\quad\kappa=0.
\label{mo:pass1}
\end{equation}
\item In the limit $\mu\gg D_R$ and $\gamma\to 0$ (active-I phase), we have
\begin{equation}
\la x^2\ra=\frac{v_0^2}{2\mu^2}\quad\text{and}\quad\kappa=-\frac{3}{2}.\\
\label{mo:act1}
\end{equation}
\item In the limit $D_R\to 0$ (active-II and passive-II phases)
\begin{equation}
\la x^2\ra=\frac{v_0^2}{2\mu(2\gamma+\mu)}\quad\text{and}\quad\kappa=\frac{3(2\gamma-3\mu)}{2(2\gamma+3\mu)}.
\label{mo:pass2}
\end{equation}
\end{itemize}

The kurtosis is always negative in the active phases. On the other hand, in the passive-II phase the kurtosis is negative in the region $1/2<\gamma/\mu<3/2$ and becomes positive for $\gamma/\mu>3/2$. Note that, zero kurtosis for $\gamma/\mu=3/2$ in the $D_R\to 0$ limit of the passive-II phase does not imply a Gaussian distribution, as is evident from \eref{marg_sc}. On the other hand, for the passive-I phase, \eref{pxdgam-gaussian} implies that kurtosis and all the other higher cumulants are zero.

\section{DRABP in $d$ dimensions}
\label{sec:ddim}
The DRABP in $d$-dimensional harmonic trap can be defined as,
\bea
\dot{\bm{r}}(t)=-\mu \bm{r}(t)+v_0\,\sigma(t)\bm{\hat{n}},
\eea
where the unit vector $\bm{\hat{n}}$ undergoes rotational diffusion on the surface of a $d$-dimensional hypersphere. 

For $D_R\to 0$, denoting $x_{\parallel}$ as the coordinate along the initial orientation and $\bm{x_{\perp}}$ as the remaining $d-1$ orthogonal coordinates, we can generalize Eq.~\eqref{eq:1drtp} to,
\begin{align}
P(x_{\parallel},\bm{x_{\perp}})=\frac{2^{1-2\nu}}{B[\nu,\nu]}\frac{1}{r_0}\left[1-\left(\frac{ x_{\parallel}}{r_0}\right)^2\right]^{\nu-1}&\Theta\left(r_0-|x_{\parallel}|\right)\cr
&\times\delta^{d-1}(\bm{x_{\perp}}),
\end{align}
where $\nu=\gamma/\mu$ and $r_0=v_0/\mu$. Assuming the distribution of the initial orientation to be isotropic  (as in Sec.~\ref{sec:dr0}), the radial distribution can be readily found as,
\bea
Q(r)=\frac{4^{1-\nu}}{B[\nu,\nu]}\frac{1}{r_0}\left[1-\left(\frac{ r}{r_0}\right)^2\right]^{\nu-1}\Theta\left(r_0-r\right),
\label{rad_dist}
\eea
where $\int_{0}^{r_0}Q(r)dr=1$. Note that, at the boundary $r=r_0$, the distribution diverges for $\nu<1$ (active-II phase) while it goes to zero for $\nu>1$ (passive-II phase). It is straightforward to obtain the marginal distribution in terms of the radial distribution as (see Appendix A of  Ref.~\cite{rtp_condensation}), 
\bea
p(x)=\frac{1}{\sqrt{\pi}}\frac{\Gamma(d/2)}{\Gamma((d-1)/2)}\int_{|x|}^{\infty}\frac{dr}{r}\left(1-\frac{x^2}{r^2}\right)^{(d-3)/2}Q(r).\cr
\label{dist_marg}
\eea
Using Eq.~\eqref{rad_dist} in Eq.~\eqref{dist_marg}, and performing the integral yields,
\bea
p(x)=\frac{1}{r_0}g_{d,\nu}(|x|/r_0),
\eea
with the scaling function,

\begin{align}
g_{d,\nu}(z)&=\frac{2\Gamma(2\nu)}{\sqrt{\pi}4^{\nu}\Gamma(\nu)}\,\Gamma\left(\frac d 2\right)\,(1-z^2)^{\nu+\frac{d-3}{2}}\,\cr
&\times {}_2\tilde{F}_1\left(\frac{d-1}{2},\nu,\nu+\frac{d-1}{2},1-z^2\right)\Theta(1-z).
\label{e:ddim}
\end{align}
Note that for the special case $d=2$, the scaling function $g_{2,\nu}(z)\equiv g_{\nu}(z)$ is obtained in \eref{marg_sc}.
Thus, the marginal distribution shows a logarithmic divergence near the origin for all dimensions $d>1$,
\bea
g_{d,\nu}(z)\sim -\log(z)+O(z^2).
\eea
It is interesting to note that for dimensions $d\geq 3$ the scaled marginal distribution in \eref{e:ddim} does not show any divergence at $z=\pm 1$.


\begin{thebibliography}{99} 

\bibitem{brown1} B. Duplantier, Progress in Mathematical Physics, {\bf 47}, Birkh\"auser Basel (2006).

\bibitem{brown2} E. Frey and K. Kroy, Ann. Phys. (Leipzig)
{\bf 14}, 20 (2005).

\bibitem{eco} J. G. Skellam, Biometrika {\bf 38}, 196 (1951).

\bibitem{comp} S. N. Majumdar, Current Science {\bf 89}, 2076 (2005).

\bibitem{finance} {\it The Random Character of Stock Market Prices}, P. H. Cootner, Ed., (MIT press, Cambridge, Massachusetts, 1964).



\bibitem{berg1} H. C. Berg, Random walks in biology, (Princeton University Press, New Jersey, 1993).

\bibitem{abp1} P. Romanczuk, M. Bar, W. Ebeling, B. Lindner, and L. Schimansky-Geier, Eur. Phys. J. Special Topics {\bf 202}, 1 (2012).

\bibitem{hydrodynamics1} M. C. Marchetti, J. F. Joanny, S. Ramaswamy, T. B. Liverpool, J. Prost, Madan Rao, and R. Aditi Simha, Rev. Mod. Phys. {\bf 85}, 1143 (2013). 





\bibitem{roadmapactive} G. Gompper, R. G Winkler, T. Speck, A. Solon, C. Nardini, F. Peruani, H. Lowen, R. Golestanian, U. Benjamin Kaupp, L. Alvarez {\it et. al.}, J. Phys.: Condens. Matter {\bf 32}, 193001 (2020).


\bibitem{Sriram} S. Ramaswamy,  J. Stat. Mech. 054002 (2017).


\bibitem{Bechinger} C. Bechinger, R. Di Leonardo, H. Lowen, C. Reichhardt, G. Volpe and G. Volpe, Rev. Mod. Phys. {\bf{88}}, 045006 (2016).

\bibitem{abp0} J. R. Howse, R. A. L. Jones, A. J. Ryan, T. Gough, R. Vafabakhsh, R. Golestanian, Phys. Rev. Lett.  ~{\bf 99}, 048102 (2007).

\bibitem{cates_abprtp} M. E. Cates and J. Tailleur, EPL {\bf 101} 20010 (2013).

\bibitem{separation_abp} J. Stenhammar, R. Wittkowski, D. Marenduzzo, and M. E. Cates, Phys. Rev. Lett. {\bf 114}, 018301 (2015).

\bibitem{abp_potoski} A. Pototsky and H. Stark,  EPL {\bf 98}, 50004 (2012).

\bibitem{franosch1} C. Kurzthaler, S. Leitmann and T. Franosch, Scientific Reports {\bf 6}, 36702 (2016).




\bibitem{rtp2008} J. Tailleur and M. E. Cates, Phys. Rev. Lett. {\bf 100}, 218103 (2008). 
 
\bibitem{abprtp_comp} A. P. Solon, M. E. Cates, J. Tailleur, Eur. Phys. J. Spec. Top. {\bf 224}, 1231 (2015).

\bibitem{pressure} A. P. Solon, Y. Fily, A. Baskaran, M. E. Cates, Y. Kafri, M. Kardar, J. Tailleur,  Nature Phys. {\bf 11}, 673 (2015). 

\bibitem{rtpddim} F. Mori, P. L. Doussal, S. N. Majumdar, and G. Schehr,  
Phys. Rev. Lett. {\bf{124}}, 090603 (2020).



\bibitem{sevilla2014} F. J. Sevilla and L. A. G. Nava, 022130 (2014).


\bibitem{abp2018} U. Basu, S. N. Majumdar, A. Rosso, G. Schehr, Phys. Rev. E {\bf{98}}, 062121 (2018).

\bibitem{rtp2d_2012} K. Martens, L. Angelani, R. Di Leonardo, and L. Bocquet, Eur. Phys. J. E {\bf 35}, 84 (2012).

\bibitem{ion1} I. Santra, U. Basu, S. Sabhapandit,  Phys. Rev. E {\bf{101}}, 062120 (2020). 


\bibitem{abp2019} U. Basu, S. N. Majumdar, A. Rosso, G. Schehr, Phys. Rev. E {\bf100}, 062116 (2019).

\bibitem{abp2020} K. Malakar, A. Das, A. Kundu, K. V. Kumar, A. Dhar, 022610 (2020).


\bibitem{1drtptrap} A. Dhar, A. Kundu, S. N. Majumdar, S. Sabhapandit, G. Schehr, 
Phys. Rev. E {\bf 99}, 032132 (2019).


\bibitem{abpexp} S. C. Takatori, R. De Dier, J. Vermant, and J. F. Brady, Nature Comm. {\bf 7}, 10694 (2016). 











\bibitem{xanthus1}Y. Wu, A. D. Kaiser, Y. Jiang and M. S. Alber,  Proc. Natl. Acad. Sci., USA {\bf 106}, 1222 (2009).

\bibitem{xanthus2}S. Thutupalli, M. Sun, F. Bunyak, K. Palaniappan and J. W. Shaevitz, J. R. Soc. Interface {\bf 12}, 20150049 (2015).

\bibitem{xanthus3}S. Leonardy, I. Bulyh, and L. S-Andersen, Mol. BioSyst. {\bf 4}, 1009 (2008).

\bibitem{xanthus4}G. Liu, A. Patch, F. Bahar, D. Yllanes, R. D. Welch, M. C. Marchetti, S. Thutupalli, and J. W. Shaevitz, Phys. Rev. Lett. {\bf 122}, 248102 (2019).


\bibitem{putida1}C. S. Harwood, K. Fosnaugh and M. Dispensa, J. Bacteriol., {\bf 171}, 4063 (1989).

\bibitem{putida2} M. Theves, J. Taktikos, V. Zaburdaev, H. Stark, and C. Beta, Biophys J. {\bf 105}, 1915 (2013).

\bibitem{marine1} J. E. Johansen, J. Pinhassi, N. Blackburn, U. L. Zweifel and A. Hagström, Aquat. Microb. Ecol. {\bf 28}, 229 (2002).

\bibitem{marine2}G. M. Barbara, J. G. Mitchell, FEMS Microbiology Ecology, {\bf 44}, 79 (2003).

\bibitem{monoperitrichous} B. L. Taylor and D. E. Koshland, J.  Bacteriol., {\bf 119}, 640 (1974).

\bibitem{drabp} I. Santra, U. Basu, S. Sabhapandit, Phys. Rev. E {\bf 104}, L012601 (2021).




 
\bibitem{grossman1} R. Gro\ss mann, F. Peruani and M. B\"ar, New J. Phys. {\bf 18}, 043009 (2016).

\bibitem{detcheverry} F. Detcheverry, Phys. Rev. E {\bf 96}, 012415 (2017).


\bibitem{Verman2020} H. R. Vutukuri, M. Lisicki, E. Lauga, and J. Vermant, Nat. Comm. {\bf 11}, 2628 (2020).



\bibitem{rtp_condensation} F. Mori, P. Le Doussal, S. N. Majumdar, and G. Schehr, Phys. Rev. E {\bf 103}, 062134 (2021).
%
%
%
%
%
%
%
%
%
%
%
%
%
%
%
%
%
%
%





\end{thebibliography}
\end{document}